\documentclass[twocolumn,floatfix,prb,aps,showpacs,superscriptaddress,longbibliography]{revtex4-2}
\usepackage{amsmath, amssymb}
\usepackage{graphicx,amsmath,amssymb,color}
\usepackage{amssymb}
\usepackage{xcolor}
\usepackage{nicefrac}
\usepackage[titletoc,title]{appendix}
\usepackage{amsmath}
\usepackage{subfigure}
\usepackage[colorlinks,bookmarks=true,citecolor=blue,linkcolor=red,urlcolor=blue]{hyperref}
\usepackage[colorlinks,bookmarks=true,citecolor=blue,linkcolor=red,urlcolor=blue]{hyperref}
\usepackage{titlesec}
\definecolor{skyblue}{RGB}{0, 200, 255}
\usepackage{pifont}
\usepackage[svgnames]{xcolor}

\makeatletter
\def\@fnsymbol#1{%
  \ifcase#1\relax 
  \or \ensuremath{\color{blue}\dagger}
  \or \ensuremath{\color{red}*}
  \or \ensuremath{\mathsection}
  \or \ensuremath{\mathparagraph}
  \else\@ctrerr\fi
}
\makeatother

\begin{document}

    \title{Signature of spin liquid state in a frustrated 3D antiferromagnet}
\author{Satish Kumar}
\thanks{equal contribution}
\affiliation{Department of Physics, Indian Institute of Technology Madras, Chennai, 600036, India}

\author{U. Jena}
\thanks{equal contribution}
\affiliation{Department of Physics, Indian Institute of Technology Madras, Chennai, 600036, India}

\author{A. Bandyopadhyay}
\affiliation{ISIS Facility, Rutherford Appleton Laboratory, Chilton, Didcot, Oxon OX11 0QX, United Kingdom}
\affiliation{Department of Physics, R. B. College (A Constituent Unit of Lalit Narayan Mithila University, Darbhanga), Dalsingsarai, Samastipur, Bihar 848114, India}

\author{G. B. G. Stenning}
\affiliation{ISIS Facility, Rutherford Appleton Laboratory, Chilton, Didcot, Oxon OX11 0QX, United Kingdom}

\author{D. T. Adroja}
\affiliation{ISIS Facility, Rutherford Appleton Laboratory, Chilton, Didcot, Oxon OX11 0QX, United Kingdom}
\affiliation{Highly Correlated Matter Research Group, Physics Department, University of Johannesburg, Auckland Park 2006, South Africa}

\author{S. Petit}
\affiliation{Laboratoire L\'eon Brillouin, CEA, CNRS, Universit\'e Paris-Saclay, CE-Saclay, F-91191 Gif-sur-Yvette, France}

\author{P. Khuntia}
\email{pkhuntia@iitm.ac.in}
\affiliation{Department of Physics, Indian Institute of Technology Madras, Chennai, 600036, India}
\affiliation{Quantum Centre of Excellence for Diamond and Emergent Materials,
Indian Institute of Technology Madras, Chennai, 600036, India}
	\date{\today}
	
	\begin{abstract}
Frustrated pyrochlore lattices in $3d$ transition-metal oxides provide an ideal platform for the experimental realization of exotic quantum states, including spin liquids with non-trivial low-energy excitations arising from the massive ground-state degeneracy of corner-sharing tetrahedral networks and the competition between emergent degrees of freedom.
Here, we focus on the synthesis and characterization of ZnCrGaO$_4$, a frustrated 3D pyrochlore-like magnet, whose intrinsic cation ordering leads to unavoidable atomic-site disorder. The Curie-Weiss fit to magnetic susceptibility in the high-temperature range yields a large Curie-Weiss temperature of $-205$ K, which suggests the presence of dominant antiferromagnetic exchange interactions ($J/k_{\rm B} \sim 55$ K) between Cr$^{3+}$ ($S=3/2$) moments. Despite the presence of strong atiferromagnetic exchange interaction, the system does not show any signature of long-range magnetic ordering down to 125 mK, as evidenced by the specific heat and $ac$ susceptibility measurements. Furthermore, the absence of bifurcation between the ZFC-FC magnetic susceptibilities measured at 0.01 T, indicates the absence of spin-freezing, which is further supported by the frequency-independent response observed in the $ac$ magnetic susceptibility measurements down to 250 mK. The presence of broad maxima in the magnetic specific heat and ac susceptibility at low temperatures suggests the development of short-range spin correlations in the dynamic state.
A power-law behavior observed in the specific heat below 1 K suggests the presence of exotic low-energy excitations and algebraic spin correlations. Our work provides compelling signatures of a dynamic ground state and unconventional low-energy excitations in a highly frustrated pyrochlore, highlighting as potential avenue for studying $S>1/2$ frustrated 3D magnets that are emerging contenders to host spin liquids.

\end{abstract}

\maketitle
   
    
Frustrated quantum materials composed of corner-sharing triangles and tetrahedra—including kagome, hyperkagome, and pyrochlore networks—as well as systems with exchange frustration mediated by bond-dependent anisotropic interactions, such as the Kitaev honeycomb lattice. These quantum materials provide an ideal platform for investigating quantum many-body phenomena such as the long-sought quantum spin liquid (QSL) arising from the complex interplay between competing degrees of freedom and frustration induced strong quantum fluctuations, promising to address some of the fundamental questions in quantum condensed matter physics, while also offering potential applications in emerging quantum technologies~\cite{anderson1973resonating, khatua2023experimental,broholm2020quantum,arh2022ising,khuntia2019novel,PhysRevB.106.104404,PhysRevB.109.024427,PhysRevB.92.180411}.
QSLs are highly entangled states of quantum matter characterized by the absence of symmetry-breaking phase transitions down to absolute zero temperature, exotic fractionalized excitations, and emergent gauge fields~\cite{khuntia2020gapless,broholm2020quantum,castelnovo2008magnetic}. However, only a few three-dimensional quantum spin-liquid candidates have been identified to date, making such systems exceptionally rare and of considerable current interest.
For instance, a compelling realization of a 3D QSL is provided by the Cu$^{2+}$ ($S$ = 1/2) based 3D hyperkagome PbCuTe$_2$O$_6$. It shows the defining characteristics of a QSL~\cite{PhysRevLett.116.107203}, such as signatures of short-range correlations, dispersive continua in the magnetic spectrum, and itinerant fractionalized excitations associated with a spinon Fermi surface~\cite{PhysRevB.90.035141,PhysRevLett.116.107203,chillal2020evidence,PhysRevLett.131.256701}. Interestingly, another promising $S=1/2$ three-dimensional bipartite lattice system exhibits quantum spin liquid behavior with a gapless excitation spectrum, stabilized by competing exchange interactions and magnetic anisotropy~\cite{gomilvsek2026quantum}. In this context, frustrated spinels hosting pyrochlore-like highly frustrated spin lattices provide a rich platform for the emergence of highly degenerate low-energy manifolds, where the intricate interplay between competing degrees of freedom stabilizes a myriads of exotic quantum phases~\cite{tsurkan2021complexity}. In particular, pyrochlore magnets, in which magnetic ions occupy a three-dimensional network of corner-sharing tetrahedra, represent one of the most fertile settings for emergent phenomena in condensed matter physics~\cite{lee_emergent_2002,castelnovo2008magnetic}. The Heisenberg antiferromagnet on a frustrated pyrochlore lattice with nearest-neighbor interactions is predicted to host a quantum-disordered state with exotic low-energy spin dynamics and gauge, even in the $S>1/2$ spin systems.~\cite{PhysRev.102.1008,PhysRevLett.80.2933,PhysRevX.9.011005}.

In particular, chromium-based spinel oxides, ACr$_2$O$_4$ (A = Zn, Mg), are prototypical examples for investigating magnetic frustration in a 3D pyrochlore Heisenberg antiferromagnetic lattice~\cite{RevModPhys.82.53}. In these materials, Cr$^{3+}$ ($S=3/2$) ions form an ideal pyrochlore lattice with a short first-nearest-neighbor bond, which results in strong magnetic frustration with a large Curie-Weiss temperature ($\theta_\text{CW}$). Despite the high degree of frustration with $\theta_\text{CW}=-390$ K, ZnCr$_2$O$_4$ undergoes long-range ordering at low temperature with $T_\text{N}=12.5$ K~\cite{PhysRevLett.84.3718}. This ordering is driven not by a conventional magnetic instability, but by strong spin-lattice coupling that lifts the degeneracy of the frustrated manifold through a symmetry-lowering structural distortion~\cite{PhysRevLett.103.037201}. A tetragonal distortion occurs as a result of a first-order magnetostructural transition at 12.5 K, which relieves frustration via a spin-driven Jahn-Teller effect reminiscent of a spin-Peierls instability in 1D spin chain~\cite{PhysRevLett.84.3718,PhysRevLett.94.137202,PhysRevLett.134.086702}. Another Cr$^{3+}$-based frustrated system is the breathing pyrochlore LiGaCr$_4$O$_8$ with a large $\theta_\text{CW}=-658$ K, which exhibits long-range ordering at $T_\text{N}=13.8$ K~\cite{PhysRevLett.110.097203}. The commonality among all these Cr$^{3+}$-based pyrochlores lies in the prominent feature of spin excitations in their ordered states. A strong dispersionless resonance mode emerges in the ordered state, which originates from the collective motions of antiferromagnetic hexagonal spin clusters whose directors constitute local zero-energy modes~\cite{lee2002emergent,PhysRevLett.101.177401,PhysRevLett.110.077205,PhysRevLett.127.147205}. Here, the spins are strongly correlated within the hexagonal cluster, while the coupling between adjacent loops remains weak.   

\begin{figure}[t]
		\begin{center}
			\includegraphics[height=195.79901pt, width=250.6988pt]{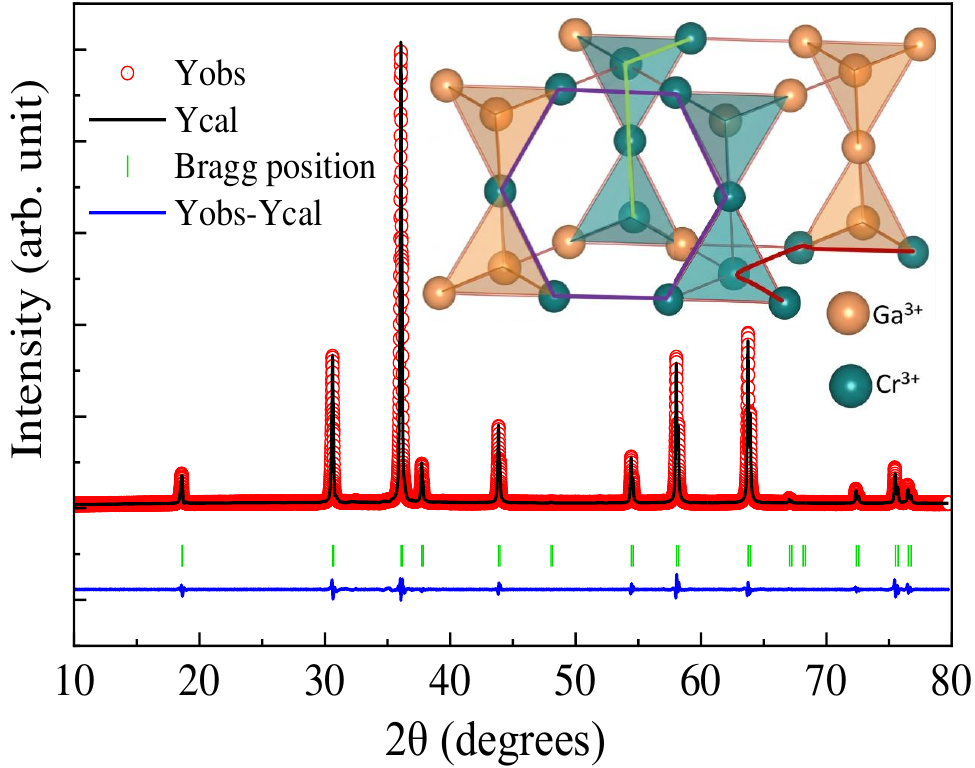}
			\caption{(a) The Rietveld refinement profile of the powder X-ray diffraction pattern taken at room temperature. The 
            inset depicts a 3D network of corner-sharing tetrahedra formed by Cr$^{3+}$ ions, with unavoidable 50$\%$ site sharing of Cr$^{3+}$ with Ga$^{3+}$ ions. The inset also illustrates an aperiodic distribution of hexagonal spin loops in $\text{ZnCrGaO}_4$. Unlike the periodic hexagon configurations found in pure $\text{ZnCr}_2\text{O}_4$, the unavoidable site sharing of nonmagnetic $\text{Ga}^{3+}$ ions at Cr$^{3+}$ ion site destroys long-range structural periodicity, leaving a disordered array of intact, local hexagonal clusters (color-coded to indicate localized loop orientations).
}
			\label{Fig.1}
		\end{center}
	\end{figure}

Another prototypical 3D QSL candidate is the garnet Gd$_3$Ga$_5$O$_{12}$, in which neutron diffuse scattering and reverse Monte Carlo analysis revealed highly correlated short-range spin textures extending beyond nearest neighbors~\cite{paddison2015hidden}. In the spin-liquid regime, the correlated spins organize into emergent ten-spin loop structures characterized by director-like multipolar correlations rather than conventional dipolar order~\cite{paddison2015hidden}. 
Correlated disorder, either inherent to the underlying frustrated spin lattice or introduced via controlled substitutions, is not merely a perturbation to the spin system; rather, it can generate highly degenerate manifolds, strong spin fluctuations, and robust short-range spin correlations, thereby offering an alternative route for stabilizing exotic quantum many-body states, including spin-liquids, in correlated quantum matter~\cite{keen2015crystallography,osborn2025diffuse}. For instance, the $S=1$ based pyrochlore NaCaNi$_2$F$_7$ displays a broad continuum of fractionalized excitations and a Coulomb-like phase evidenced by low-energy pinch points in neutron scattering experiments.~\cite{plumb2019continuum}.
Similarly, in another 3D magnet, CsNiCrF$_6$, Ni$^{2+}$ and Cr$^{3+}$ cations together form a pyrochlore lattice and self-organize into fully packed, same-species loops composed exclusively of either Ni$^{2+}$ or Cr$^{3+}$ moments~\cite{fennell2019multiple,xvxt-whns}. The lengths of these loops follows a power-law distribution, and the magnetic dynamics are governed by collective excitations associated with these loops rather than individual spins~\cite{xvxt-whns}. 
The experimental realization of QSLs in higher-dimensional spin lattices, particularly in three dimensions, remains a daunting challenge, as the increased connectivity in the spin Hamiltonian and reduced quantum fluctuations often lead to spin freezing or long-range magnetic ordering.  In this context, promising three-dimensional spin lattices with correlated disorder offer a viable route for stabilizing a QSL state while preserving emergent looplike excitations, as predicted in three-dimensional pyrochlore magnets~\cite{broholm2020quantum}.

\begin{figure*}
	\centering
	\includegraphics[width=\textwidth]{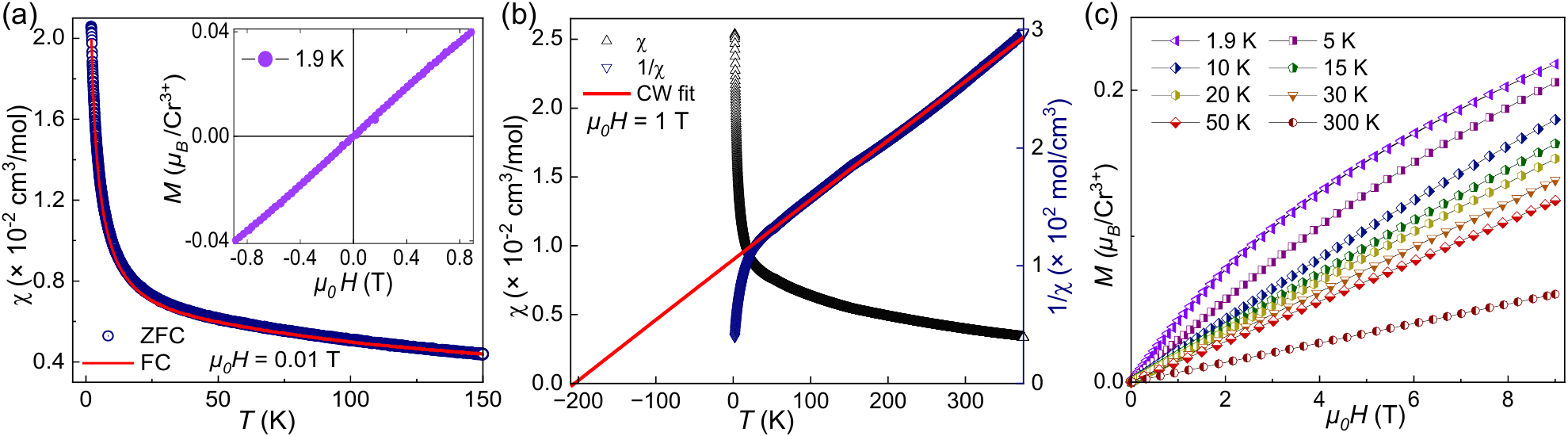}
	\caption{(a) Zero-field-cooled and field-cooled magnetic susceptibility as a function of temperature under an applied field of 0.01 T. The inset displays the $M$ vs $H$ in five quadrants measured at 1.9 K.
    (b) Temperature dependence of the magnetic susceptibility ($\chi$) (Left, Y-Axis) and inverse susceptibility ($\chi$) (Right, Y-Axis) measured at an applied field of 1 T. The solid red line represents a Curie–Weiss fit to the high-temperature data.
    (c) Magnetization as a function of external magnetic field at several temperatures.}
    \label{Fig:ZCGO_Mag}
\end{figure*}

Here, we present the synthesis, structural details and thermodynamic results on a frustrated 3D magnet ZnCrGaO$_4$ (ZCGO). 
Despite the strong antiferromagnetic exchange interaction with $\theta_\text{CW}=-205$ K ( $J/k_{\rm B}=55$ K), the material does not undergo a magnetic phase transition down to 125 mK, as evidenced by specific heat and $ac$ susceptibility measurements. In addition, the frequency-independent nature of the $ac$ susceptibility down to 225 mK rules out spin freezing in spite of the unavoidable site sharing between Cr$^{3+}$ and Ga$^{3+}$ ions in this 3D magnet. A broad maximum observed in the zero-field magnetic specific heat ($C_{\text m}$) data around 12.5 K implies the onset of short-range spin correlations, which is further supported by the change of 37\%  of the magnetic entropy expected for the Cr$^{3+}$ ($S=3/2$) system. A power law behavior of magnetic specific heat, $C_{\text m}\sim T^2$ below 1 K signals the presence of a low-energy gapless excitation spectrum. Our results demonstrate that unavoidable correlated disorder and strong frustration conspire to evade both long-range ordering and spin-glass behavior, and the material hosts spin liquid-like behavior down to 125 mK. This behavior identifies ZCGO as a compelling realization of a critical cooperative paramagnet, where the system resides near a highly degenerate manifold and sustains persistent spin dynamics with algebraic spin correlations.


Polycrystalline samples of ZCGO have been prepared by the standard solid state reaction route~\cite{SM}. The Rietveld refinement of the XRD measurement taken at room temperature is shown in Fig.~\ref{Fig.1}.
Structural analysis confirms that ZCGO crystallizes in the cubic spinel structure with the space group \textit{Fd}$\bar{3}$\textit{m}, and a refined lattice parameter of $a = 8.252$ \AA. This structure is isostructural to mixed-cation spinels such as ZnAlGaO$_4$ and ZnFeGaO$_4$, in which the pyrochlore sublattice is occupied in a 1:1 ratio by Al$^{3+}$/Ga$^{3+}$ and Fe$^{3+}$/Ga$^{3+}$ ions, respectively~\cite{maayouf1993x,verger2016spectroscopic}. Random substitution of nonmagnetic Ga$^{3+}$ ions on the half of the Cr sites fragments the corner-sharing tetrahedral network into a scale-free distribution of closed Cr$^{3+}$ loops and finite clusters. In the parent compound ZnCr$_2$O$_4$, the degenerate ground state can be mapped via a highly ordered, periodic assignment of spins into non-overlapping hexagonal loops~\cite{lee_emergent_2002}. A fundamentally different scenario emerges in ZCGO due to the half-doped Ga$^{3+}$ dilution at Cr$^{3+}$ site as confirmed by x-ray absorption spectroscopy~\cite{SM}. The substitution of nonmagnetic Ga$^{3+}$ ions at Cr$^{3+}$ site breaks the continuous exchange pathways of the pyrochlore network. Consequently, any global, periodic tiling of hexagonal loops is rendered impossible. Instead, the arrangement of these hexagonal spin loops becomes strictly aperiodic as shown in the inset of Fig.~\ref{Fig.1}.
Unlike the CsNiCrF$_6$ pyrochlore, where loops are fully packed and composed of two different species, ZCGO hosts a single-species loop ensemble whose collective dynamics govern the low-energy behavior. 

\begin{figure*}
	\centering
	\includegraphics[width=\textwidth]{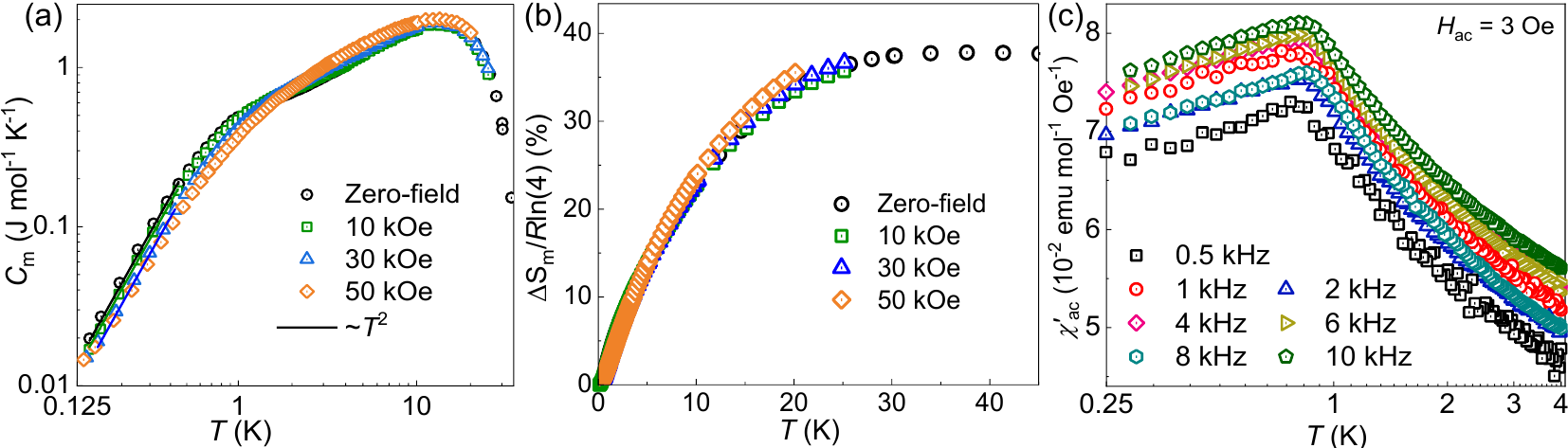}
	\caption{
    (a) The temperature dependence of the magnetic specific heat ($C_m$) after subtracting the lattice part from the total specific heat. The solid lines depicts a fit to $C_m \sim T^2$. Full details of lattice subtraction and DE-fit are given in Supplementary Material~\cite{SM}. 
    (b) The magnetic entropy changes $\Delta S_{\rm m}$, which is around 37\% of the expected value. i.e. Rln(4) for one Cr$^{3+}$ ion per formula unit   
    (c) The real part of the $ac$ magnetic susceptibility $\chi'_{\text{ac}}$ of ZCGO measured at various frequencies under an excitation field of $H_{\text{ac}}$ = 3 Oe.
    }
    \label{Fig:ZCGO_HC_AC}
\end{figure*}

The $dc$ magnetic susceptibility measured under both zero-field-cooled (ZFC) and field-cooled (FC) conditions in an applied magnetic field of $\mu_0 H=0.01$ T shows no bifurcation down to 2 K, suggesting the absence of spin freezing. [see Fig.~\ref{Fig:ZCGO_Mag}(a)]. Furthermore, the five-quadrant field-sweep isothermal magnetization measured at 1.9 K provides additional evidence for the absence of spin freezing, ruling out the presence of parasitic phases, grain-boundary effects, or defect-induced magnetic contributions in the sample [inset of Fig.~\ref{Fig:ZCGO_Mag}(a)]. In order to gain insight into the magnetic interaction between the moments of Cr$^{3+}$ (\textit{S} = 3/2) ions, the inverse magnetic susceptibility ($1/\chi$) recorded in an applied field of 1 T was fitted with the Curie-Weiss (CW) law $\chi= \chi_{0} + C/(T-\theta_{\rm CW}$). Here, $\theta_{\rm CW}$ is the Curie-Weiss temperature, $\chi_{0}$ represents the temperature-independent magnetic susceptibility arising from core diamagnetic and  Van Vleck contributions,  and $C$ is the Curie constant which is related to the effective moment by $\mu_{\rm eff} = \sqrt{8C}~\mu_{B}$. The CW fit [Fig.~\ref{Fig:ZCGO_Mag}(b)] in the temperature range 100 K $\leq T \leq$ 375 K yields $\theta_{\rm CW} = -205\pm1$ K, $C = 1.938\pm0.005$ cm$^3$·K·mol$^{-1}$, with a temperature-independent susceptibility $\chi_{0} \approx 5.67\times10^{-5}$ cm$^3$/mol. The obtained effective moment, $\mu_\text{eff} = 3.94 \mu_\text{B}$, agrees well with the spin-only value of $3.87 \mu_\text{B}$ expected for free Cr$^{3+}$ ($S=3/2$) ions. A comparatively large $\theta_{\rm CW}$, indicating dominant antiferromagnetic interactions between the Cr$^{3+}$ moments, and its reduced value compared to the parent compound ZnCr$_2$O$_4$ (with $\theta_\text{CW}=- 390$ K)~\cite{takagi_highly_2011}, suggests a weakening of the antiferromagnetic exchange interaction due to unavoidable site disorder. As shown in Fig.~\ref{Fig:ZCGO_Mag} (b), a clear change in the slope of $1/\chi$ below 50 K marks the onset of antiferromagnetic spin correlations, where spins begin to form locally entangled clusters driven by the interplay of exchange randomness and frustration~\cite{ji_spin-lattice_2009}. Fig.~\ref{Fig:ZCGO_Mag}(c) shows the field-dependent magnetization $M(H)$ observed at various temperatures in the range of 1.9 K $\leq T \leq$ 300 K, which yields no saturation up to 9 T, and is consistent with dominant antiferromanetic interactions.

The exchange strength $J$ between the nearest neighbors obtained from the Curie-Weiss temperature, is given by $J/k_\text{B}=-\frac{3\theta_\text{CW}}{z_{\text{eff}}S(S+1)}$ as per the mean-field approximation, where $k_\text{B}$ is the Boltzmann constant, and z denotes the coordination number of each Cr$^{3+}$ moments, for an ideal 3D pyrochlore lattice of corner-sharing tetrahedra, $z=6$~\cite{kittel_introduction_2018}. However, due to site disorder, the coordination number is reduced to $z_{\text{eff}}$ = 3, yielding an average exchange strength of $J/k_{\rm B} \approx$ 55 K. The spin stiffness ($\rho_s$) in the large $S$ limit, quantifies the energy cost associated with spin distortions, as $\rho_s \sim \frac{JS^2}{d}$, where $d$ is the nearest-neighbor distance between Cr$^{3+}$ ions, which is 2.917 (\AA)~\cite{kittel_quantum_1963, auerbach_interacting_2012, PhysRevB.48.13170}. Using this relation, the $\rho_s$ is calculated as $\sim$ 6.09 meV/\AA, indicating robust antiferromagnetic exchange interactions.
 We also calculate the theoretical upper limit for the propagation speed (c) of conventional magnetic excitations, which is given by c = $\frac{2\sqrt{D}JSd}{\hbar}$. For a three-dimensional system ($D$ = 3), this calculation yields a propagation speed of $\sim 1.09\times10^4$ m/sec~\cite{southern_spin_1993}. 
 Despite having a magnetic-site occupancy above the pyrochlore lattice percolation threshold ($p_c \sim 0.39$)~\cite{PhysRev.133.A310, PhysRevLett.84.3450}, the system's spin stiffness is significantly softened by disorder and follows the relation as $\rho_s(p) \sim \rho_{s,0} (p - p_c)^t$, where $t$ is the critical exponent that governs how rapidly the stiffness vanishes as the magnetic site occupancy \(p\) approaches \(p_{c}\) ~\cite{RevModPhys.45.574,stauffer_introduction_2018}. The combined effect of this reduced stiffness and strong geometric frustration suppresses long-range magnetic order in ZCGO.

Specific heat is a powerful tool for tracking magnetic correlations and characterizing the density of states in a frustrated pyrochlore, where subtle ground states compete. In order to scrutinize the ground state of this frustrated 3D magnet, specific heat measurements were conducted down to 125 mK in magnetic fields for 0 $\leq$ $\mu_oH$ $\leq$ 9 T [see Fig.~\ref{Fig:ZCGO_HC_AC}(a)], revealing the absence of any $\lambda$-like anomaly, which would otherwise signify the onset of long-range magnetic order. The absence of magnetic ordering is consistent with strong magnetic frustration, as quantified by the large frustration parameter, \( f = \frac{|\theta_\text{CW}|}{T_{\text{N}}}\sim 1640 \). Such a large value indicates strongly enhanced frustration-induced spin fluctuations, and the presence of a macroscopically degenerate ground state prevents the formation of conventional long-range order.

The magnetic specific heat data, plotted as $C_m$ vs $T$ in a logarithmic scale~[see Fig.~\ref{Fig:ZCGO_HC_AC}(a)], reveal a broad maximum around 12.5 K in zero and higher fields. This feature is characteristic of the emergence of short-range spin correlations, a hallmark of geometrically frustrated magnets, including QSL candidates~\cite{bastien-spin-2019}. 
The magnetic specific heat, $C_{m}$, follows a power-law dependence ($C_{\rm m} \sim  T^2 $) below 1~K, as shown in Fig.~\ref{Fig:ZCGO_HC_AC}(a), indicative of exotic low-energy excitations and unconventional spin dynamics characteristic of frustrated magnets. This behavior is consistent with a spin-liquid state, where algebraic spin correlations emerge within a highly degenerate manifold of low-energy excitations~\cite{lester_magnetic-field-controlled_2021,khatua2023experimental}. The corresponding magnetic entropy change is about 37\% [see Fig.~\ref{Fig:ZCGO_HC_AC}(b)] of that expected for a Cr$^{\rm 3+}$ ($S = 3/2$) system, adding further credence to the short-range magnetic correlation scenario, as evidenced by the magnetic specific heat.

\begin{figure}
    \centering
    \includegraphics[width=1\linewidth]{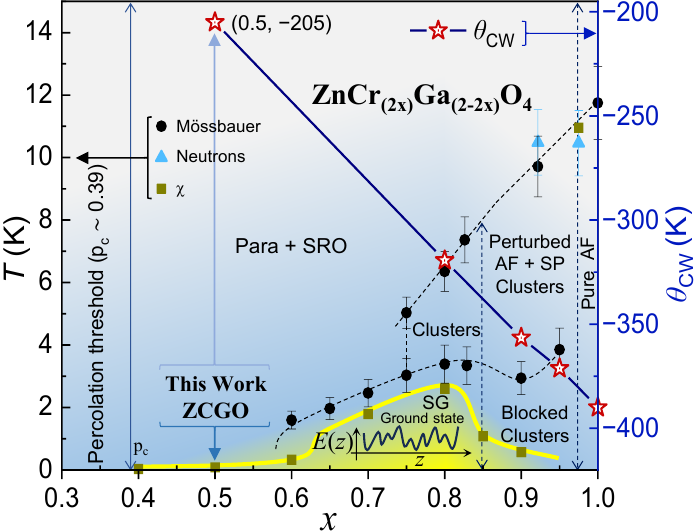}
    \caption{Magnetic phase diagram of frustrated pyrochlore, ZnCr$_{(2x)}$Ga$_{(2-2x)}$O$_4$ depicting exotic phases. The phase diagram is determined by Mössbauer spectra [solid-circle \ding{108}], neutron diffraction [solid-triangle \textcolor{skyblue}{\ding{115}}], and magnetic susceptibility ($\chi$) [solid-square \textcolor{olive}{\ding{110}}]. A spin-glass phase has been observed in the region 0.6$\leq$ x $\leq0.85$, highlighted in yellow~\cite{fiorani1983magnetic,saifi_mossbauer_1988,PhysRevB.77.014405}. The star-line (-\textcolor{Red}{\ding{75}}-) represents the behavior of the $\theta_\text{CW}$ at the right y-axis depending upon the chemical pressure ($x$).
     The inset shows a typical shape of rugged free-energy \textit{E}(\textit{z}) landscapes for spin glass ground state, where \textit{z} represents the different spin configuration~\cite{jena2026topological}. 
     }\label{Fig4:ZCGO}
\end{figure}

The $ac$ magnetic susceptibility is a sensitive probe for identifying slow magnetic relaxation and spin-glass behavior. By applying a weak ac magnetic field, the in-phase ($\chi'$) and out-of-phase ($\chi''$) components of the magnetic response can be separated, corresponding to the reversible and dissipative parts of the magnetization dynamics, respectively. The $ac$ magnetic susceptibility measurements were carried out down to 250 mK under an applied $ac$ magnetic field of $H_{\mathrm{ac}} = 3$ Oe over the frequency range of 500 Hz--10 kHz [see Fig.~\ref{Fig:ZCGO_HC_AC}(c)]. The real part, $\chi'(T)$, shows a broad maximum near 0.8 K with no observable frequency dependence within the experimental resolution, in contrast to the characteristic frequency-dependent peak expected for a canonical spin glass. The broad peak in $ac$ susceptibility is a hallmark of short-range magnetic correlations, which is in agreement with the specific heat results, further supporting a spin-liquid state in ZCGO~\cite{PhysRevB.106.104404,TbBO3_bx3h-4g62}. Previous studies have shown that a genuine spin-glass phase is observed only within the compositional range $0.6 \leq x \leq 0.85$ for ZnCr$_{2x}$Ga$_{2-2x}$O$_4$~\cite{fiorani1983magnetic,saifi_mossbauer_1988}, as reflected in the phase diagram presented in Fig.~\ref{Fig4:ZCGO}.
Interestingly, despite unavoidable correlated site disorder~\cite{keen2015crystallography}, the highly frustrated pyrochlore spin lattice
stabilizes a dynamic liquid-like ground state. Further confirmation of this scenario requires detailed $\mu$SR and/or neutron-scattering measurements down to the sub-Kelvin regime; however, this is beyond the scope of the present study, although it provides an impetus for future investigations.

Our results reveal that ZnCrGaO$_4$  belongs to the broader family of Cr-based spinel antiferromagnets and highlight its exceptional degree of magnetic frustration on a pyrochlore lattice. Canonical spinels such as ZnCr$_2$O$_4$ and MgCr$_2$O$_4$ undergo antiferromagnetic ordering at $T_{\mathrm N} \approx 12$ K despite strong antiferromagnetic interactions ($\theta_{\mathrm{CW}} \sim -400$ K), yielding frustration parameters of $f\sim 30$. In contrast, the present material with unavoidable correlated disorder in ZCGO stabilizes a disordered ground state (with $\theta_{\mathrm{CW}} = -205$~K, corresponding to $f = 1640$, which implies exceptionally strong frustration among all known Cr spinel antiferromagnets) down to 125~mK. This remarkable enhancement of $f$ in ZCGO demonstrates that the combined effects of geometric frustration intrinsic to the pyrochlore lattice and intrinsic but correlated disorder owing to inevitable nonmagnetic Ga$^{3+}$ ions at Cr$^{3+}$ site suppress the spin-lattice instability that ordinarily relieves frustration in Cr spinels. Consequently, ZCGO emerges as an extreme limit of the Cr-spinel family, where strong antiferromagnetic correlations persist without conventional magnetic ordering. These results establish ZCGO as a compelling platform for investigating unconventional cooperative phenomena, disorder-driven criticality, and spin-liquid-like behavior.

$\text{ZnCrGaO}_4$ resides at a critical dilution limit ($x = 0.5$) where the intrinsic site sharing of $\text{Ga}^{3+}$ with $\text{Cr}^{3+}$ ions drives the system towards a dynamic spin state.
The $ac$ susceptibility measurements demonstrate no frequency dependence down to 250 mK, indicating that ZCGO remains free from spin freezing or glassiness. Despite a large Curie–Weiss temperature ($\theta_\text{CW}$), the heat capacity does not exhibit sharp anomalies indicative of symmetry-breaking order and instead shows power-law behavior reminiscent of an algebraic spin liquid. In the parent compounds ZnCr$_2$O$_4$ and MgCr$_2$O$_4$, the magnetic correlations eventually condense into spin-lattice ordered states through a spin-Peierls instability~\cite{PhysRevLett.134.086702}. Previous neutron-scattering studies on related Cr spinels have shown that persistent spin correlations are naturally described in terms of collective hexagonal loop modes~\cite{lee2002emergent,PhysRevLett.101.177401,xvxt-whns,w6j5-ljfj,PhysRevB.77.014405}, which constitute the fundamental local degrees of freedom of the pyrochlore lattice. The lack of a thermodynamic phase transition in ZCGO suggests that it evades the spin-Peierls-like magneto-structural transition characteristic of the parent compound $\text{ZnCr}_2\text{O}_4$, pointing instead toward a dynamic spin liquid state. In this context, ZCGO represents a uniquely engineered structural route to preserve local hexagonal loop excitations while simultaneously evading both long-range Néel ordering and spin freezing.


To summarize, inherent correlated disorder resulting from the unavoidable site-sharing of Cr$^{3+}$ and Ga$^{3+}$ ions in the pyrochlore lattice gives rise to bond randomness. 
This leads to spatially inhomogeneous magnetic correlations, and frustration-induced strong spin fluctuations prevent conventional long-range magnetic ordering down to sub-Kelvin temperatures, which is almost three orders of magnitude lower than the characteristic interaction energy scale. The parent compound
ZnCr$_2$O$_4$ exhibits a structural phase transition
near the antiferromagnetic ordering temperature associated
with spin-Peierls’ instability; however, the ZCGO appears
to be stable even at milli-Kelvin temperatures. Despite the inherent site disorder above the percolation threshold on a pyrochlore lattice, the system shows no signatures of spin freezing. Specific heat measurements reveal the onset of antiferromagnetic short-range spin correlations around $\sim 12.5$ K. The power-law behavior of zero-field magnetic specific heat suggests the realization of spin liquid with exotic low-energy excitations and algebraic spin correlations in this Cr-pyrochlore. A highly degenerate low-energy manifold and short-range spin correlations driven by correlated disorder in a strongly frustrated pyrochlore lattice stabilize a spin-liquid state with algebraic spin correlations, highlighting three-dimensional quantum magnets as a promising avenue for the experimental realization of exotic quantum phases.

\noindent\hspace{1em} \textit{Acknowledgements.}
 P.K. acknowledges the funding by the Anusandhan National Research Foundation (ANRF), Department of Science and Technology, India through  Research Grants.

\bibliography{ZCGO}

\end{document}